\documentclass[12pt]{article}
\usepackage{graphicx}
\usepackage{cite}

\newcommand\pubnumber{SI-HEP/14-18, QFET/14-12}
\newcommand\pubdate{01 August 2014}

\def\siegen{Theoretische Physik 1, Naturwissenschaftlich-Technische Fakult\"at, \\
Universit\"at Siegen, Walter-Flex-Stra\ss{}e 3, D-57068 Siegen, Germany}
\def\support{\footnote{Supported in parts by the Bundesministerium f\"ur Bildung und Forschung (BMBF),
and by the Deutsche Forschungsgemeinschaft (DFG) within Research Unit FOR 1873
(``Quark Flavour Physics and Effective Field Theories'').}}

\def\Title#1{\begin{center} {\Large #1 } \end{center}}
\def\Author#1{\begin{center}{ \sc #1} \end{center}}
\def\Address#1{\begin{center}{ \it #1} \end{center}}

\newcommand\pubblock{\rightline{\begin{tabular}{l} \pubnumber\\
         \pubdate  \end{tabular}}}
\newenvironment{Abstract}{\begin{quotation}  }{\end{quotation}}
\newenvironment{Presented}{\begin{quotation} \begin{center} 
             PRESENTED AT\end{center}\bigskip 
      \begin{center}\begin{large}}{\end{large}\end{center} \end{quotation}}
\def\Acknowledgements{\bigskip  \bigskip \begin{center} \begin{large}
             \bf ACKNOWLEDGEMENTS \end{large}\end{center}}
\def\beq{\begin{equation}}
\def\eeq#1{\label{#1}\end{equation}}
\def\eeqn{\end{equation}}

\def\beqa{\begin{eqnarray}}
\def\eeqa#1{\label{#1}\end{eqnarray}}
\def\eeqan{\end{eqnarray}}

\let\bar=\overbar

\def\Dslash{\not{\hbox{\kern-4pt $D$}}}
\def\dslash{\not{\hbox{\kern-2pt $\del$}}}

\def\msb{{\bar{\ssstyle M \kern -1pt S}}}

\begin{document}
\begin{titlepage}
\pubblock

\vfill
\Title{Non-Leptonic Heavy Meson Decays -- Theory Status}
\vfill
\Author{Thorsten Feldmann\support}
\Address{\siegen}
\vfill
\begin{Abstract}
I briefly review the status and recent progress in the theoretical
understanding of non-leptonic decays of beauty and charm hadrons.
Focusing on a personal selection of topics, this covers 
perturbative calculations in quantum chromodynamics,
analyses using flavour symmetries of strong interactions,
and the modelling of the relevant hadronic input functions.
\end{Abstract}
\vfill
\begin{Presented}
Flavor Physics and CP Violation (FPCP-2014),\\  Marseille, France, May 26-30 2014
\end{Presented}
\vfill
\end{titlepage}
\def\thefootnote{\fnsymbol{footnote}}
\setcounter{footnote}{0}

\section{Motivation}

%\index 
%\cite 

Non-leptonic decays of hadrons containing a heavy
bottom or charm quark may provide important information
on the angles of the Cabibbo-Kobayashi-Maskawa (CKM)
matrix in the Standard Model (SM). They also may
reveal deviations from the SM, in particular the 
presence of new CP-violating phases from 
``new physics'' (NP). 
The non-trivial hadronic dynamics
in such flavour transitions further allows to assess the accuracy
of theoretical methods (perturbative or non-perturbative) 
in Quantum Chromodynamics (QCD),
based on the separation (``factorization'') of short-
and long-distance strong-interaction effects.
Finally, comparison of experimental data and theoretical parametrizations
may also lead to a better understanding of 
the hadronic structure of heavy-light bound states.

Experimental studies of 
non-leptonic $B$- and $D$-meson decays, which have been successfully
carried out at flavour factories and at hadron colliders in the past,
will be continued at present and future experiments, notably at
LHCb and Belle-II.
With the foreseen increasing experimental
precision, theoretical calculations should therefore catch up in
accuracy in order to achieve reliable phenomenological
conclusions about the validity of the SM or hints for NP 
(for comprehensive reviews, see e.g.\ 
\cite{Buchalla:2008jp,Antonelli:2009ws,Bediaga:2012py,Bevan:2014iga}).

In this proceedings contribution, I highlight some recent 
theoretical results and developments 
which contribute to improving
our understanding of exclusive heavy-meson 
%(and also heavy-baryon)
decays.

\section{Perturbative Calculations}

Theory predictions for non-leptonic exclusive decays, by definition, 
depend on hadronic matrix element which cannot be calculated in QCD
perturbation theory. Still, the presence of a heavy quark mass
(notably for the $b$-quark) implies that certain dynamical effects
are related to short-distance physics (on length scales of the
order $1/m_b$) and may thus be accessible in
perturbative QCD. The challenge is then to systematically 
separate short- and long-distance phenomena, where the latter
should be described by as few independent hadronic parameters
as possible.

\subsubsection*{QCD Factorization for $B$ Decays into Light Mesons}

$B$-decays into two (energetic) light mesons $(M_1,M_2)$ are
described by hadronic matrix elements of weak transition operators ${\cal O}_i$. 
In the limit of infinitely heavy quark masses, the strong-interaction
dynamics  factorizes
according to \cite{Beneke:1999br}
\begin{eqnarray}
 \langle M_1 M_2 | {\cal O}_i | B\rangle 
 &\simeq& F^{B \to M_1} \, \int du \, {T_i^{\rm I}(u)} \, \phi_{M_2}(u) \cr 
  && + \int d\omega \, du \, dv \, {T_i^{\rm II}(\omega,u,v)} \, \phi_B(\omega) \,
    \phi_{M_1}(v) \, \phi_{M_2}(u) \,.
    \label{qcdf}
\end{eqnarray} 
Here, the hadronic input functions
are given by (universal) transition form factors $F^{B \to M}$ evaluated at large recoil
energy,
and light-cone distribution amplitudes (LCDAs) $\phi_{B,M_1,M_2}$ for heavy
and light mesons which depend on the momenta (respectively momentum fractions)
of the light quarks.
The short-distance kernels $T_i^{\rm I,II}$ can be calculated perturbatively,
including renormalization-group (RG) improvement in the framework of
soft-collinear effective theory (SCET \cite{Bauer:2000yr,Beneke:2002ph}). 
One generic phenomenological 
consequence of this ``QCD-improved'' factorization (QCDF) is that direct CP violation
in these decays -- which requires strong rescattering phases -- is suppressed
by ${\cal O}(\alpha_s)$ and/or ${\cal O}(1/m_b)$ in the heavy-quark limit.

The result of QCDF calculations can be parametrized 
in terms of 
(decay-channel dependent) tree- and penguin-amplitude parameters $\alpha_i$,
for instance,\footnote{Here $\lambda_q$ denote combinations of CKM elements,
and the normalization factors $A_{M_1 M_2}$ are given in terms of
form factors and decay constants \cite{Beneke:2003zv}.}
\begin{eqnarray}
 \langle \pi^+\pi^-| {\cal H}_{\rm eff} |\bar B^0\rangle 
 &=& A_{\pi\pi} \big\{ \lambda_u \left[{\alpha_1(\pi\pi)} + 
 {\alpha_4^u(\pi\pi)} \right]
 + \lambda_c \, {\alpha_4^c(\pi\pi)} \big\} \cr 
% -\langle \pi^0\pi^0| {\cal H}_{\rm eff} |\bar B^0\rangle 
% &=& A_{\pi\pi} \big\{ \lambda_u \left[{\alpha_2(\pi\pi)} - 
% {\alpha_4^u(\pi\pi)} \right]
% - \lambda_c \, {\alpha_4^c(\pi\pi)} \big\} \cr
 \langle \pi^+\bar K^-| {\cal H}_{\rm eff}|\bar B^0\rangle &=&
  A_{\pi \bar K} \big\{ \lambda_u^{(s)} \left[ 
  {\alpha_1(\pi K)} + {\alpha_4^u(\pi K)}\right]
  + \lambda_c^{(s)} 
   \, {\alpha_4^c(\pi K)} \big\}
   \cr 
   & \mbox{etc.}
\end{eqnarray}
The current status of higher-order calculations (NNLO, i.e.\
second order in the strong coupling $\alpha_s$) 
is as follows.
\begin{itemize}
 \item 2-loop vertex corrections contributing to $T_i^{\rm I}$
  for tree-amplitude parameters $\alpha_{1,2}$ have been determined 
  independently in \cite{Bell:2007tv,Beneke:2009ek}.
  The corresponding 2-loop vertex corrections for penguin amplitudes
  are currently under study \cite{BBHL}, and preliminary results will be
  shown below.
  
 \item The 1-loop spectator corrections to $T_i^{\rm II}$ 
  for tree amplitudes 
  have been analyzed in \cite{Beneke:2005vv,Kivel:2006xc,Pilipp:2007mg}.
   The 1-loop spectator corrections for penguin amplitudes can
  be found in \cite{Beneke:2006mk,Jain:2007dy}.
  
\end{itemize}
Let us briefly discuss 
the numerical significance of the individual contributions.
Considering e.g.\ the colour-allowed (-suppressed) 
tree amplitude $\alpha_1$($\alpha_2$) in $B \to \pi\pi$
decays, we have
\begin{eqnarray}
  \alpha_1(\pi\pi) &=& \phantom{-} \left[\, 1.008\,\right]_{V_0} + \left[\,0.022 + 0.009 i\,\right]_{V_1}
   + {\left[\,0.024 + 0.026 i\,\right]_{V_2}} 
   \cr 
   & \phantom{=} & \,\, - [\,0.014\,]_{S_1} - { [\,0.016 + 0.012 i\,]_{S_2}} 
   - [\,0.008\,]_{1/m_b}
   \cr 
   &=& \phantom{-} 1.015^{+0.020}_{-0.029} + \left( 0.023^{+0.015}_{-0.015} \right) i \,,
\end{eqnarray}
and
\begin{eqnarray}
\alpha_2(\pi\pi) &=& \phantom{-} \left[\, 0.224\,\right]_{V_0} - \left[\,0.174 + 0.075 i\,\right]_{V_1}
   - {\left[\,0.029 + 0.046 i\,\right]_{V_2}} 
   \cr 
   & \phantom{=} & \,\, + [\,0.084\,]_{S_1} + { [\,0.037 + 0.022 i\,]_{S_2}} 
   + [\,0.052\,]_{1/m_b}
   \cr 
   &=& \phantom{-} 0.194^{+0.130}_{-0.095} - \left( 0.099^{+0.057}_{-0.056} \right) i \,,
 \end{eqnarray}
where $V_{0,1,2}$ stand for (LO,NLO,NNLO) contributions to $T_i^{\rm I}$,
while (NLO,NNLO) spectator contributions from $T_i^{\rm II}$ are labeled by
$S_{1,2}$. 
Estimates of $1/m_b$ power corrections are also quoted. One observes that
\begin{itemize}
 \item the perturbative expansion is well behaved, with 
  individual {NNLO} corrections $V_2$ and $S_2$ 
  being significant but tending to cancel in the sum;
 \item precise predictions are achieved for the colour-allowed tree amplitude $\alpha_1$,
  while larger hadronic uncertainties remain for the colour-suppressed amplitude $\alpha_2$;
 \item the relative phase between $\alpha_1$ and $\alpha_2$ stays small at NNLO.
\end{itemize}
Penguin amplitudes are currently known at NLO, and for the $\pi\pi$ channel one gets
\begin{eqnarray}
 \alpha_4^u(\pi\pi) &=& -0.024^{+0.004}_{-0.002} + \left(-0.012^{+0.003}_{-0.002} \right) i \cr 
 \alpha_4^c(\pi\pi) &=& -0.028^{+0.005}_{-0.003} + \left(-0.006^{+0.003}_{-0.002} \right) i
\end{eqnarray}
The calculation of penguin amplitudes at NNLO, which is currently worked out
\cite{BBHL}, involves ${\cal O}(70)$ 2-loop diagrams with up to 3 independent mass scales
($m_b$, $m_c$, $u m_b$) and 4 external legs and non-trivial charm thresholds at $ (1-u) \, m_b^2 = 4m_c^2$
 (where $u$ is the momentum fraction of a quark in a light meson).
 Preliminary results have been shown in recent conference talks 
 \cite{Talks}.

In the past, the QCDF results have been used for 
comprehensive phenomenological studies: 
\begin{itemize} 
\item The decays $B \to \pi\pi,\pi\rho,\rho\rho$ have been investigated in 
\cite{Beneke:2009ek,Bell:2009fm}. Predictions for colour-suppressed modes 
are rather uncertain and typically underestimated in QCDF
(depending on the hadronic matrix element governing the size of spectator-scattering
contributions). The uncertainties from hadronic form factors and the CKM element
$|V_{ub}|$ can be reduced by considering ratios with the semi-leptonic rates.
   
\item Estimates for  tree-dominated $B_s$ decays
can be found in \cite{BellCKM2010}. Here the relevant hadronic parameters
(form fators and LCDAs) are less well known than for the previous case.
On the other hand, the pattern of annihilation contributions is simpler.
It has also been emphasized that the size of charming-penguin effects
can be tested from ratios of colour-allowed modes \cite{Zhu:2010eq}.

\item One can also find results for charmless $B$-meson decays
into scalar mesons \cite{Cheng:2013fba}.

\end{itemize}

\begin{figure}[t!!!bph]
 \begin{center}
\includegraphics[width=0.6\textwidth]{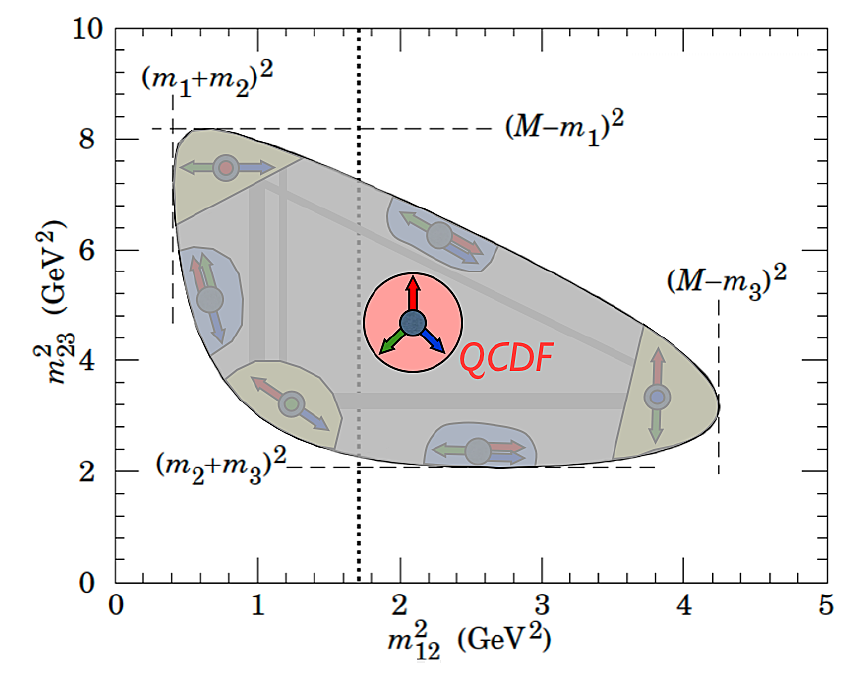}
\end{center}
\caption{Dalitz plot of $B\to \pi\pi\pi$ decays, highlighting the 
``mercedes-star'' configuration where QCDF is applicable (from \cite{HKMRV}).
\label{fig:3pi}}
\end{figure}

\subsubsection*{Other modes / Further Activities}

NNLO corrections in QCDF for the decays $B \to D \pi$
are currently studied as well. In this case, the charm and bottom quark
are usually treated as heavy quarks in HQET, and their hard fluctuations are 
integrated out at a common matching scale. Compared to charmless decays, 
this leads to a number of new master integrals
which depend on the mass ratio $m_c^2/m_b^2$.
The status of the computation has been recently reported in
\cite{Huber:2014kaa}.

The systematics of QCDF in e.g.\ $B \to \pi\pi$ decays can also be independently 
addressed by phenomenological
studies of the related  decay $B \to \pi\pi \ell \nu$. 
In the 
region where the dipion invariant mass is large,
$m_{\pi\pi}^2 \sim {\cal O}(m_b^2)$, one obtains a factorization theorem \cite{DvDTF:2014}
which takes a similar form as in $B \to \pi\pi$ (with $\otimes$ denoting the convolution
integrals as in (\ref{qcdf})),
\begin{eqnarray}
 \langle \pi^+\pi^-|\bar u \, \Gamma \, b |B\rangle 
& \simeq &  F^{B \to \pi} \cdot \phi_\pi \otimes {T_\Gamma^{\rm I}(q^2)}
 \ + \ \phi_\pi \otimes \phi_\pi \otimes \phi_B(\omega) \otimes {T_\Gamma^{\rm II}(q^2)} \,.
\end{eqnarray}
In contrast to $B \to \pi\pi$, the decay is now induced by 
semi-leptonic operators, where 
 $\Gamma=\gamma_\mu(1-\gamma_5)$ in the SM.
In the considered kinematic region, at least one hard gluon
is required to produce the additional back-to-back quark-antiquark
pair in the final state. As a consequence, the kernel $T_\Gamma^{\rm I}$
starts at ${\cal O}(\alpha_s)$, while the kernel $T_\Gamma^{\rm II}$ 
includes additional spectator interactions and thus starts at ${\cal O}(\alpha_s^2)$.
Moreover, the variable $q^2$ representing the invariant mass  of the lepton
pair, provides a new lever arm to assess systematic uncertainties related 
to non-factorizable effects.

Finally, let me  mention that the QCDF approach can also be applied 
to 3-body decays $B \to \pi\pi\pi$ in the kinematic region where each
individual dipion mass $m_{ij}^2 =(p_i+p_j)^2$ is sufficiently large,
i.e.\ the three momenta forming a ``mercedes-star''-like configuration,
see Fig.~\ref{fig:3pi}.
A systematic theoretical and phenomenological investigation of this
idea will be pursued in \cite{HKMRV}.

\section{Flavour Symmetries in QCD}

The approximate flavour symmetries (FS) of light quarks in
strong interaction dynamics (isospin for $u,d$ quarks,
$U$-spin for $d,s$ quarks, or the full $SU(3)_F$ for $u,d,s$)
have always been a standard
tool in understanding hadronic physics. The wealth of 
experimental data on non-leptonic $b$- and $c$-decays nowadays 
allows to draw conclusions about first-order FS-breaking 
corrections. In combination with factorization approaches,
this can be used to test assumptions about subleading terms in the $1/m_b$
expansion, and in the long run this may also 
enhance the sensitivity to finding deviations from the
SM in these decay modes.

\subsubsection*{Isospin and $SU(3)_F$ in $B \to PP$ and $B\to PV$}

The complete set of isospin, $U$-spin and $SU(3)_F$ relations
among the CP asymmetries in $B$-meson decays to two pseudoscalars
(or to one pseudoscalar and one vector meson),
together with first-order symmetry-breaking effects,
has recently been analyzed in \cite{Grossman:2013lya} (see also \cite{He:2013vta,Gronau:2013mda}).
Comparing with experimental data, the amount of
$SU(3)_F$ breaking turns out to be of reasonable size, e.g.\
\begin{eqnarray} \tilde\Delta & \equiv & \frac{\delta_{\rm CP}[B_d \to K^+\pi^-]+\delta_{\rm CP}[B_s \to K^-\pi^+]}{
  \delta_{\rm CP}[B_d \to K^+\pi^-]-\delta_{\rm CP}[B_s \to K^-\pi^+]} = 0.026 \pm 0.106 \,,
\end{eqnarray}
if the observables are properly normalized, 
$\delta_{\rm CP}[i\to f] = \frac{8\pi\, m_i^2}{|\vec p_{i\to f}|} \, \Delta_{\rm CP}[i\to f]$.
Additional constraints on $SU(3)_F$ breaking can be obtained from
certain theory approaches. As an example, the authors of \cite{Grossman:2013lya} compare 
two phenomenological approaches based on different treatment of non-factorizable effects.
In the so-called ``BBNS'' approach \cite{Beneke:2003zv}, the combination of
amplitude parameters $\alpha_1 \alpha_4^c$ determines the numerically dominant source of
strong phases and direct CP violation.
In the ``BPRS'' approach \cite{Bauer:2004tj} the dominant source of non-factorizable effects
is expected from  charm-penguin contributions. This leads to different
correlations between $SU(3)_F$ breaking in BRs and CP asymmetries for 
($B_s \to K^-\pi^+$, $B_d\to \pi^-K^+$) and ($B_d\to\pi^-\pi^+$, $B_s \to K^-K^+$) 
which can be confronted with experimental
data. Present data are still consistent with both alternatives, but with improved 
precision in future experiments more decisive conclusions about 
non-factorizable $SU(3)_F$-breaking effects in non-leptonic decays will be possible.

\begin{figure}[t!!bph]
\begin{center}
 \includegraphics[width=0.55\textwidth]{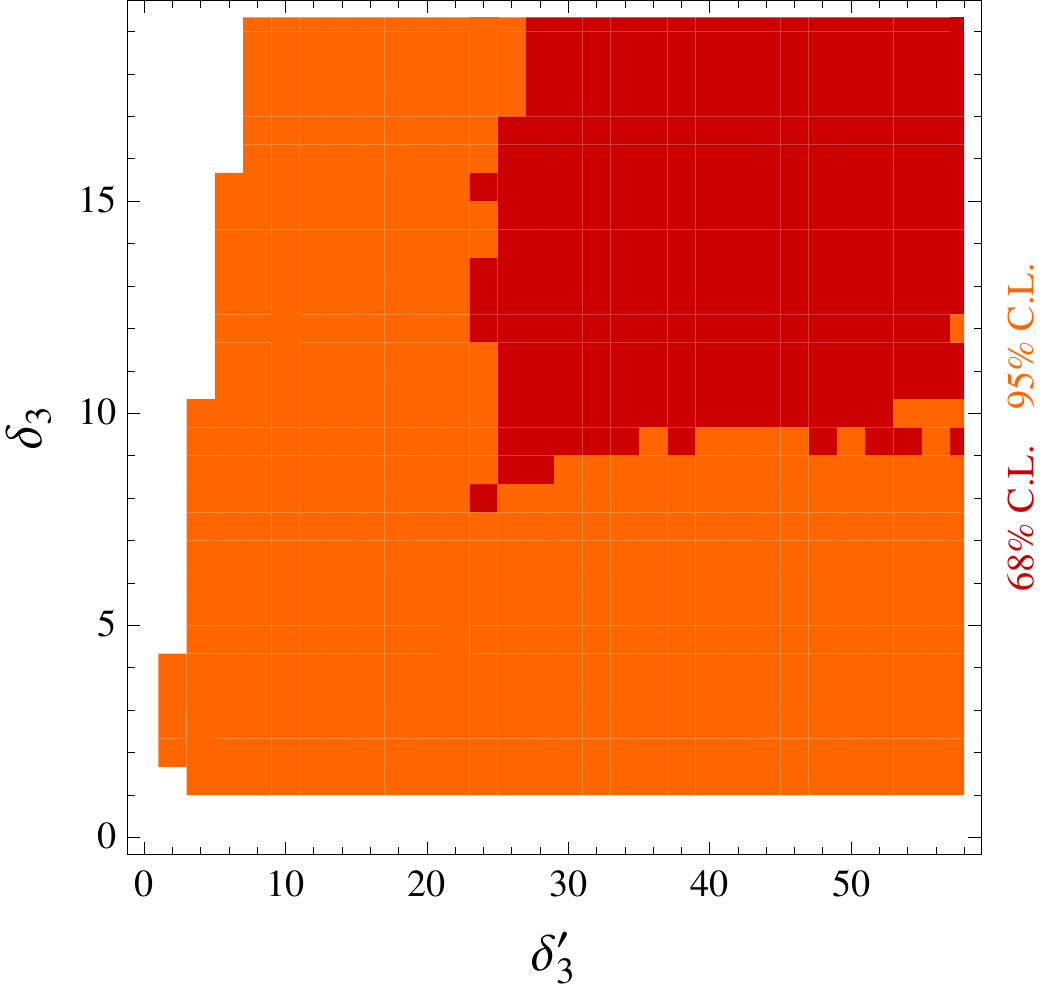}
\end{center}
\caption{\label{figMJ2} Fit of penguin-enhancement parameters $\delta_3,\delta_3'$ 
to experimental data on non-leptonic charm decays (after {\sc Charm2013}).
Orange (red) regions indicate the 95\% (68\%) C.L.\ regions,
see \cite{Hiller:2012xm} for more details.
Fig.\ taken from \cite{Hiller:2012xm}.}
\end{figure}

\subsubsection*{Non-leptonic Charm Decays}

Factorization-based approaches to describe exclusive
charm decays suffer from the relatively 
small charm-quark mass, such that non-factorizable
corrections are more important than for the corresponding
bottom decays. Again, approximate flavour symmetries
of QCD turn out to be helpful.
The flavour anatomy of $D$-meson decays into two pseudoscalars 
has been systemetically studied in \cite{Hiller:2012xm}.
Taking into account the complete set of first order $SU(3)_F$-breaking effects,
a consistent fit of the available experimental data could be achieved,
with corrections to the flavour-symmetry limit of natural size, 
${\cal O}(30\%)$. 
At the time of the analysis, the measured value of the difference between CP asymmetries 
in $D^0 \to K^+K^-$ and $D^0 \to \pi^+\pi^-$ decays, $\Delta a_{\rm CP}^{\rm dir}(K^+K^-,\pi^+\pi^-)$,
required a drastic penguin enhancement (in the SM), where 
also $a_{\rm CP}^{\rm dir}(D^0 \to K_SK_S)$, 
$a_{\rm CP}^{\rm dir}(D_s \to K_S \pi^+)$, and  $a_{\rm CP}^{\rm dir}(D_s \to K^+\pi^0)$
contribute. After the {\sc Charm2013} conference, the effect is less dramatic, as 
illustrated in Fig.~\ref{figMJ2}.
Alternatively, the $SU(3)_F$ analysis of various CP asymmetries in 
non-leptonic $D$-meson decays allows to discriminate between
different NP scenarios.

A related $SU(3)_F$ analysis of non-leptonic $D$-meson
decays to two pseudoscalars or one pseudoscalar and one vector
meson can be found in \cite{Grossman:2012ry}.
The focus of that work is to derive sum rules among
decay amplitudes or decay rates such that ${\cal O}(m_s)$ 
effects drop out. For instance, from the CKM-weighted
amplitude relations 
\begin{eqnarray}
 |(D^0|\pi^+\rho^-)|/\lambda + |(D^0|K^+K^{*-})|/\lambda
& = & |(D^0|K^+\rho^-)|/\lambda^2 + |(D^0|\pi^+K^{*-})
\end{eqnarray} 
one can infer a prediction for one of the yet unmeasured 
branching ratios (BRs),
\begin{eqnarray}
 & \Rightarrow & {\rm Br}(D^0 \to \rho^- K^+) \simeq (1.7\pm 0.4) \cdot 10^{-4} \,.
\end{eqnarray}
The formalism can also be combined with the 
``$\Delta U=0$ rule'' for large penguins \cite{Brod:2012ud}
to predict direct CP asymmetries in $D \to PV$ decays.
 
\subsubsection*{Other modes}

Recent analyses based on $SU(3)_F$ symmetry have also
been performed for particular final state configurations in $B$-meson
decays to three light pseudoscalars \cite{Bhattacharya:2014eca};
and for 2-body $B$-meson decays into octet or decuplet baryons \cite{Chua:2013zga}.
In a systematic study of $SU(3)$-breaking effects in $B \to J/\psi P$ decays 
\cite{Jung:2012mp} it has been shown how penguin corrections
      can be extracted from data, constraining the pollution 
      in the extraction of the CKM angle $\sin2\beta$
      to be very small, 
      $|\Delta S|\leq 0.01$. With more precise data on CP asymmetries 
      and BRs this uncertainty can be further reduced in the future.

\section{Hadronic Input Functions}

Besides the perturbative computation of short-distance kernels in factorization theorems,
and factorization-independent constraints from flavour symmetries, an important role for 
the quantitative prediction of nonleptonic decays is played by universal hadronic input
parameters like decay constants, form factors and light-cone distribution amplitudes (LCDAs).
As these contain the information about hadronic binding effects, they have to be determined 
by nonperturbative methods (i.e.\ lattice or sum rules) or extracted from experimental data.

\subsubsection*{Transition Form Factors}

If factorization in (\ref{qcdf}) holds, the  form factors for $B \to M_1$ transitions 
together with the decay constants $f_{M_2}$ determine the overall magnitude of $B \to M_1 M_2$
decay amplitudes and BRs at leading order. The sensitivity to this
type of hadronic input can be reduced by considering \emph{ratios} with semileptonic 
decays or among different nonleptonic decays. 
For theoretical estimates of transition form factors between heavy and light mesons, it 
has become customary to perform combined fits of light-cone sum-rule results 
(valid at large energy transfer, see e.g.\ \cite{Ball:2004ye,Duplancic:2008ix})
and results from QCD simulations on the lattice (valid at low recoil energy,
see the discussion in \cite{latticetalk} and references given therein)
on the basis of the so-called ``$z$-expansion''.

\subsubsection*{Light-Cone Distribution Amplitudes for the $B$-Meson}

The $B$-meson LCDA $\phi_B(\omega)$ determines the size of the spectator interactions in 
QCDF. The leading term in the short-distance kernels $T_i^{\rm II}$ 
is proportional to the inverse moment  
\begin{eqnarray}
 \lambda_B^{-1}(\mu) & \equiv & \int_0^\infty \frac{d\omega}{\omega} \, \phi_B(\omega,\mu) \,,
\end{eqnarray}
where $\omega$ denotes the light-cone projection of the spectator quark's energy.
The size of this parameter is crucial for phenomenological analyses. While 
comparison with data on $B \to \pi\pi,\pi\rho,\rho\rho$ decays prefers relatively
small values around $\lambda_B \sim 200$~MeV within the QCDF/BBNS approach, 
QCD-sum-rule based estimates typically lead to values
in the region $\lambda_B \sim (350-500)$~MeV \cite{Braun:2003wx}. 
Recent OPE analyses of the large-scale behaviour of $\phi_B$, using the 
concept of ``dual'' LCDAs \cite{Bell:2013tfa} (see also \cite{Braun:2014vba}),
have shown that -- contrary to naive expectation --
the parameter $\lambda_B$ is essentially independent of other HQET parameters
like $\bar\Lambda= M_B - m_b$ \cite{Feldmann:2014ika}.
The most promising approach to independently determine $\lambda_B$ is to extract its value
from experimental data on $B \to \gamma\ell\nu$ decays, on the basis of 
QCD factorization theorems and estimates for non-factorizable $1/m_b$ corrections 
\cite{Beneke:2011nf,Braun:2012kp}. Presently, using the BaBar bound for that decay
rate from 2009, one finds $\lambda_B > 115$~MeV.

\subsubsection*{Annihilation Parameters in QCDF}

Amplitude topologies where the spectator quark in a $B$-meson 
annihilates with the $b$-quark via weak interactions also play
a crucial role for the phenomenological analyses of non-leptonic
$B$ decays. Notably, these decay topologies cannot be
described by heavy-to-light form factors. The $1/m_b$
power corrections induced by annihilation topologies lead to
IR-sensitive convolution integrals in QCDF and can thus only
be modelled in a crude and ad-hoc manner. In a recent 
phenomenological analysis of pure annihilation decays of $B_d$ and 
$B_s$ mesons \cite{Wang:2013fya} (see also \cite{Chang:2012xv})
the flavour dependence of annihilation parameters in QCDF has
been studied. Comparing, on the one hand, $B_d \to \pi^- K^+$ 
and $B_s \to \pi^+ K^-$ decays, one expects similar strong
rescattering phases because the final states are related 
by charge conjugation, which turns out to be in line with
experimental observation. On the other hand, comparing 
$B_s \to \pi^+\pi^-$ and $B_d \to K^+ K^-$ within that approach, sizeable 
$SU(3)_F$ breaking effects are required. 

The size of strong phases in non-leptonic $B$-meson decays 
has also been studied within a phenomenological rescattering 
model \cite{Gronau:2012gs}. Distinguishing different topological
amplitudes (``exchange (E)'', ``annihilation (A)'', ``penguin-annihilation (PA)''),
experimental data reveals  a relatively regular pattern, 
where (E)~$\sim \ (5-10)\%$ and 
      (PA)~$\sim (15-20)\%$ 
of the largest amplitude from which they can rescatter.
This allows one to estimate several BRs for 
not yet observed $B$ and $B_s$ decays.

%%%%%%%%%%%%%%%%%%%%%%%%%%%%%%%%%%%%%%%%%%%%%%%%%%%%%%%%%%%%%%%%%%%%%%%%%
%%
%%   use this format to include an .eps figure into your paper
%%
% \begin{figure}[htb]
% \centering
% %\includegraphics[height=1.5in]{magnet}
% \caption{Plan of the magnet used in the mesmeric studies.}
% \label{fig:magnet}
% \end{figure}
%%%%%%%%%%%%%%%%%%%%%%%%%%%%%%%%%%%%%%%%%%%%%%%%%%%%%%%%%%%%%%%%%%%%%%%%%%%

% 
% %%%%%%%%%%%%%%%%%%%%%%%%%%%%%%%%%%%%%%%%%%%%%%%%%%%%%%%%%%%%%%%%%%%%%%%%%
% %%
% %%   use this format to include a LaTeX table  into your paper
% %%
% \begin{table}[t]
% \begin{center}
% \begin{tabular}{l|ccc}  
% Patient &  Initial level($\mu$g/cc) &  w. Magnet &  
% w. Magnet and Sound \\ \hline
%  Guglielmo B.  &   0.12     &     0.10      &     0.001  \\
%  Ferrando di N. &  0.15     &     0.11      &  $< 0.0005$ \\ \hline
% \end{tabular}
% \caption{Blood cyanide levels for the two patients.}
% \label{tab:blood}
% \end{center}
% \end{table}
% %%%%%%%%%%%%%%%%%%%%%%%%%%%%%%%%%%%%%%%%%%%%%%%%%%%%%%%%%%%%%%%%%%%%%%%%%%%

\section{Summary/Outlook}

The dynamics of strong interactions in non-leptonic 
decays of heavy mesons 
%(and baryons) 
is extremely complex. While one has to admit that on the theory side 
a conceptual breakthrough for the systematic calculation of 
non-factorizable hadronic effects is still lacking,
the combination of several theoretical methods in many cases still 
gives a satisfactory phenomenological picture.
\begin{itemize}
 \item Short-distance kernels in the QCD factorization approach 
   are now being calculated at NNLO for a variety of decays.
 \item Systematic studies of 
      $SU(3)_F$ flavour-symmetry breaking effects on the basis of phenomenological 
      data are available.
 \item The ongoing improvement of the experimental situation leads to 
       better knowledge on hadronic input parameters and more reliable estimates 
       of systematic theoretical uncertainties. 
\end{itemize}
We are thus looking forward to phenomenological updates that 
combine the state-of-the-art results for radiative corrections,
hadronic input parameters, and  $SU(3)_F$-breaking effects.

\Acknowledgements
I would like to thank the organizers of FPCP'14 for the invitation and 
for a stimulating workshop program. I am also grateful to Guido Bell,
Tobias Huber and Martin Jung for valuable scientific input.


\begin{thebibliography}{99}

%%
%%  bibliographic items can be constructed using the LaTeX format in SPIRES:
%%    see    http://www.slac.stanford.edu/spires/hep/latex.html
%%  SPIRES will also supply the CITATION line information; please include it.
%%

%\cite{Buchalla:2008jp}
\bibitem{Buchalla:2008jp}
  M.~Artuso, D.~M.~Asner, P.~Ball, E.~Baracchini, G.~Bell, M.~Beneke, J.~Berryhill and A.~Bevan {\it et al.},
  %``$B$, $D$ and $K$ decays,''
  Eur.\ Phys.\ J.\ C {\bf 57} (2008) 309
  [arXiv:0801.1833 [hep-ph]].
  %%CITATION = ARXIV:0801.1833;%%
  %203 citations counted in INSPIRE as of 26 Jun 2014

%\cite{Antonelli:2009ws}
\bibitem{Antonelli:2009ws}
  M.~Antonelli, D.~M.~Asner, D.~A.~Bauer, T.~G.~Becher, M.~Beneke, A.~J.~Bevan, M.~Blanke and C.~Bloise {\it et al.},
  %``Flavor Physics in the Quark Sector,''
  Phys.\ Rept.\  {\bf 494} (2010) 197
  [arXiv:0907.5386 [hep-ph]].
  %%CITATION = ARXIV:0907.5386;%%
  %191 citations counted in INSPIRE as of 26 Jun 2014

%\cite{Bediaga:2012py}
\bibitem{Bediaga:2012py}
  RAaij {\it et al.}  [LHCb Collaboration],
  %``Implications of LHCb measurements and future prospects,''
  Eur.\ Phys.\ J.\ C {\bf 73} (2013) 2373
  [arXiv:1208.3355 [hep-ex]].
  %%CITATION = ARXIV:1208.3355;%%
  %102 citations counted in INSPIRE as of 26 Jun 2014

%\cite{Bevan:2014iga}
\bibitem{Bevan:2014iga}
  A.~J.~Bevan, B.~Golob, T.~Mannel, S.~Prell, B.~D.~Yabsley, K.~Abe, H.~Aihara and F.~Anulli {\it et al.},
  %``The Physics of the B Factories,''
  arXiv:1406.6311 [hep-ex].
  %%CITATION = ARXIV:1406.6311;%%

  
%%%%%%%%%%%%%%%%%%%%%%%%%%%%%%%%%%%%%%%%%%%%%%%%%%%%%%%%%%%%%%%%%%%%%%%%%%%%%%%%%%%%%%%%%%%%%%%%%%%%%%%%%%%%  
  
%\cite{Beneke:1999br}
\bibitem{Beneke:1999br}
  M.~Beneke, G.~Buchalla, M.~Neubert and C.~T.~Sachrajda,
  %``QCD factorization for B ---> pi pi decays: Strong phases and CP violation in the heavy quark limit,''
  Phys.\ Rev.\ Lett.\  {\bf 83} (1999) 1914
  [hep-ph/9905312].
  %%CITATION = HEP-PH/9905312;%%
  %1030 citations counted in INSPIRE as of 02 Jun 2014
%
%\cite{Beneke:2000ry}
%\bibitem{Beneke:2000ry}
%  M.~Beneke, G.~Buchalla, M.~Neubert and C.~T.~Sachrajda,
  %``QCD factorization for exclusive, nonleptonic B meson decays: General arguments and the case of heavy light final states,''
  Nucl.\ Phys.\ B {\bf 591} (2000) 313
  [hep-ph/0006124];
  %%CITATION = HEP-PH/0006124;%%
  %952 citations counted in INSPIRE as of 02 Jun 2014
  %
%\cite{Beneke:2001ev}
%\bibitem{Beneke:2001ev}
%  M.~Beneke, G.~Buchalla, M.~Neubert and C.~T.~Sachrajda,
 %``QCD factorization in B ---> pi K, pi pi decays and extraction of Wolfenstein parameters,''
  Nucl.\ Phys.\ B {\bf 606} (2001) 245
  [hep-ph/0104110];
  %%CITATION = HEP-PH/0104110;%%
  %830 citations counted in INSPIRE as of 02 Jun 2014 

%%%%%%%%%%%%%%%%%%%%%%%%%%%%%%%%%%%%%%%%%%%%%%%%%%%%%%%%%%%%%%%%%%%%%%%%%%%%%%%%%%%%5  

  
%\cite{Bauer:2000yr}
\bibitem{Bauer:2000yr}
  C.~W.~Bauer, S.~Fleming, D.~Pirjol and I.~W.~Stewart,
  %``An Effective field theory for collinear and soft gluons: Heavy to light decays,''
  Phys.\ Rev.\ D {\bf 63} (2001) 114020
  [hep-ph/0011336].
  %%CITATION = HEP-PH/0011336;%%
  %778 citations counted in INSPIRE as of 26 Jun 2014
  
  %\cite{Beneke:2002ph}
\bibitem{Beneke:2002ph}
  M.~Beneke, A.~P.~Chapovsky, M.~Diehl and T.~Feldmann,
  %``Soft collinear effective theory and heavy to light currents beyond leading power,''
  Nucl.\ Phys.\ B {\bf 643} (2002) 431
  [hep-ph/0206152].
  %%CITATION = HEP-PH/0206152;%%
  %345 citations counted in INSPIRE as of 26 Jun 2014

  
%\cite{Beneke:2003zv}
\bibitem{Beneke:2003zv}
  M.~Beneke and M.~Neubert,
  %``QCD factorization for B ---> PP and B ---> PV decays,''
  Nucl.\ Phys.\ B {\bf 675} (2003) 333
  [hep-ph/0308039].
  %%CITATION = HEP-PH/0308039;%%
  %738 citations counted in INSPIRE as of 10 Jul 2014
  
%%%%%%%%%%%%%%%%%%%%%%%%%%%%%%%%%%%%%%%%%%%%%%%%%%%%%%%%%%%%%%%%%%%%%%%%%%%%%%%%%%%%%%%


  
%\cite{Bell:2007tv}
\bibitem{Bell:2007tv}
  G.~Bell,
  %``NNLO vertex corrections in charmless hadronic B decays: Imaginary part,''
  Nucl.\ Phys.\ B {\bf 795} (2008) 1
  [arXiv:0705.3127 [hep-ph]];
  %%CITATION = ARXIV:0705.3127;%%
  %45 citations counted in INSPIRE as of 02 Jun 2014%\cite{Bell:2009nk}
%\bibitem{Bell:2009nk}
%  G.~Bell,
  %``NNLO vertex corrections in charmless hadronic B decays: Real part,''
  Nucl.\ Phys.\ B {\bf 822} (2009) 172
  [arXiv:0902.1915 [hep-ph]].
  %%CITATION = ARXIV:0902.1915;%%
  %29 citations counted in INSPIRE as of 02 Jun 2014


  %\cite{Beneke:2009ek}
\bibitem{Beneke:2009ek}
  M.~Beneke, T.~Huber and X.-Q.~Li,
  %``NNLO vertex corrections to non-leptonic B decays: Tree amplitudes,''
  Nucl.\ Phys.\ B {\bf 832} (2010) 109
  [arXiv:0911.3655 [hep-ph]].
  %%CITATION = ARXIV:0911.3655;%%
  %27 citations counted in INSPIRE as of 02 Jun 2014
  
  
  
\bibitem{BBHL}
G.~Bell, M.~Beneke, T.~Huber,and X.-Q.~Li,
in preparation.

\bibitem{Talks} 
M.~Beneke at ``3rd KEK Flavor Factory Workshop'', Tsukuba, Feb 13-15, 2014;
T.~Huber at ``Loops and Legs 2014'', Weimar, Apr 27 - May 2;
G.~Bell at ``BEACH 2014'', Birmingham, Jul 21–26, 2014.



%%%%%%%%%%%%%%%%%%%%%%%%%%%%%%%%%%%%%%%%%%%%%%%%%%%%%%%%%%%%%%%%%  
  
 %\cite{Beneke:2005vv}
\bibitem{Beneke:2005vv}
  M.~Beneke and S.~J\"ager,
  %``Spectator scattering at NLO in non-leptonic b decays: Tree amplitudes,''
  Nucl.\ Phys.\ B {\bf 751} (2006) 160
  [hep-ph/0512351].
  %%CITATION = HEP-PH/0512351;%%
  %85 citations counted in INSPIRE as of 02 Jun 2014 
  
  
  %\cite{Kivel:2006xc}
\bibitem{Kivel:2006xc}
  N.~Kivel,
  %``Radiative corrections to hard spectator scattering in B ---> pi pi decays,''
  JHEP {\bf 0705} (2007) 019
  [hep-ph/0608291].
  %%CITATION = HEP-PH/0608291;%%
  %34 citations counted in INSPIRE as of 02 Jun 2014
  
  %\cite{Pilipp:2007mg}
\bibitem{Pilipp:2007mg}
  V.~Pilipp,
  %``Hard spectator interactions in B ---> pi pi at order alpha(2)**s,''
  Nucl.\ Phys.\ B {\bf 794} (2008) 154
  [arXiv:0709.3214 [hep-ph]].
  %%CITATION = ARXIV:0709.3214;%%
  %31 citations counted in INSPIRE as of 02 Jun 2014
  
%%%%%%%%%%%%%%%%%%%%%%%%%%%%%%%%%%%%%%%%%%%%%%%%%%%%%%%%%%%%%%%%%%%%%%%%%%%  
  
  %\cite{Beneke:2006mk}
\bibitem{Beneke:2006mk}
  M.~Beneke and S.~J\"ager,
  %``Spectator scattering at NLO in non-leptonic B decays: Leading penguin amplitudes,''
  Nucl.\ Phys.\ B {\bf 768} (2007) 51
  [hep-ph/0610322].
  %%CITATION = HEP-PH/0610322;%%
  %60 citations counted in INSPIRE as of 02 Jun 2014
  
  
  
  %\cite{Jain:2007dy}
\bibitem{Jain:2007dy}
  A.~Jain, I.~Z.~Rothstein and I.~W.~Stewart,
  %``Penguin Loops for Nonleptonic B-Decays in the Standard Model: Is there a Penguin Puzzle?,''
  arXiv:0706.3399 [hep-ph].
  %%CITATION = ARXIV:0706.3399;%%
  %42 citations counted in INSPIRE as of 02 Jun 2014
  
%%%%%%%%%%%%%%%%%%%%%%%%%%%%%%%%%%%%%%%%%%%%%%%%%%%%%%%%%%%%%%%%%%%%%%%%%

  
  %\cite{Bell:2009fm}
\bibitem{Bell:2009fm}
  G.~Bell and V.~Pilipp,
  %``B- ---> pi- pi0/rho- rho0 to NNLO in QCD factorization,''
  Phys.\ Rev.\ D {\bf 80} (2009) 054024
  [arXiv:0907.1016 [hep-ph]].
  %%CITATION = ARXIV:0907.1016;%%
  %23 citations counted in INSPIRE as of 02 Jun 2014
  
  
\bibitem{BellCKM2010}  
G.~Bell, talk presented at CKM~2010, University of Warwick, Sep~2010.

%\cite{Zhu:2010eq}
\bibitem{Zhu:2010eq}
  G.~Zhu,
  %``Bd \to \pi^- K^{(*)+} and Bs \to \pi^+(\rho^+) K^- decays with QCD factorization and flavor symmetry,''
  JHEP {\bf 1005} (2010) 063
  [arXiv:1002.4518 [hep-ph]].
  %%CITATION = ARXIV:1002.4518;%%
  %3 citations counted in INSPIRE as of 02 Jun 2014
  
%\cite{Cheng:2013fba}
\bibitem{Cheng:2013fba}
  H.-Y.~Cheng, C.-K.~Chua, K.-C.~Yang and Z.-Q.~Zhang,
  %``Revisiting charmless hadronic B decays to scalar mesons,''
  Phys.\ Rev.\ D {\bf 87} (2013) 11,  114001
  [arXiv:1303.4403 [hep-ph]].
  %%CITATION = ARXIV:1303.4403;%%
  %7 citations counted in INSPIRE as of 02 Jun 2014

  %\cite{Huber:2014kaa}
\bibitem{Huber:2014kaa}
  T.~Huber and S.~Kr\"ankl,
  %``Towards NNLO corrections in B --> D pi,''
  arXiv:1405.5911 [hep-ph].
  %%CITATION = ARXIV:1405.5911;%%
  
  
\bibitem{DvDTF:2014}
T.~Feldmann, D.~van Dyk, work in progress.

\bibitem{HKMRV}
T.~Huber, S.~Kr\"ankl, T.~Mannel, D.~Rosenthal, J.~Virto,
work in progress (private communication).

  %\cite{Grossman:2013lya}
\bibitem{Grossman:2013lya}
  Y.~Grossman, Z.~Ligeti and D.~J.~Robinson,
  %``More Flavor SU(3) Tests for New Physics in CP Violating B Decays,''
  JHEP {\bf 1401} (2014) 066
  [arXiv:1308.4143 [hep-ph]].
  %%CITATION = ARXIV:1308.4143;%%

  %\cite{He:2013vta}
\bibitem{He:2013vta}
  X.-G.~He, S.-F.~Li and H.-H.~Lin,
  %``CP Violation in $B^0_s \to K^-\pi^+$, $B^0 \to K^+\pi^-$ Decays and Tests for SU(3) Flavor Symmetry Predictions,''
  JHEP {\bf 1308} (2013) 065
  [arXiv:1306.2658 [hep-ph]].
  %%CITATION = ARXIV:1306.2658;%%
  %5 citations counted in INSPIRE as of 03 Jun 2014

  %\cite{Gronau:2013mda}
\bibitem{Gronau:2013mda}
  M.~Gronau,
  %``U-spin breaking in CP asymmetries in B decays,''
  Phys.\ Lett.\ B {\bf 727} (2013) 136
  [arXiv:1308.3448 [hep-ph]].
  %%CITATION = ARXIV:1308.3448;%%
  %10 citations counted in INSPIRE as of 03 Jun 2014
  
  
%\cite{Bauer:2004tj}
\bibitem{Bauer:2004tj}
  C.~W.~Bauer, D.~Pirjol, I.~Z.~Rothstein and I.~W.~Stewart,
  %``B ---> M(1) M(2): Factorization, charming penguins, strong phases, and polarization,''
  Phys.\ Rev.\ D {\bf 70} (2004) 054015
  [hep-ph/0401188].
  %%CITATION = HEP-PH/0401188;%%
  %313 citations counted in INSPIRE as of 10 Jul 2014
%   
  
  %\cite{Hiller:2012xm}
\bibitem{Hiller:2012xm}
  G.~Hiller, M.~Jung and S.~Schacht,
  %``SU(3)-Flavor Anatomy of Non-Leptonic Charm Decays,''
  Phys.\ Rev.\ D {\bf 87} (2013) 014024
  [arXiv:1211.3734 [hep-ph]];
  %%CITATION = ARXIV:1211.3734;%%
  %13 citations counted in INSPIRE as of 03 Jun 2014
%\cite{Hiller:2013awa}
%\bibitem{Hiller:2013awa}
%  G.~Hiller, M.~Jung and S.~Schacht,
  %``SU(3)$_{F}$ in nonleptonic charm decays,''
  PoS EPS {\bf -HEP2013} (2014) 371
  [arXiv:1311.3883 [hep-ph]].
  %%CITATION = ARXIV:1311.3883;%%
  %1 citations counted in INSPIRE as of 13 Jul 2014

  %\cite{Grossman:2012ry}
\bibitem{Grossman:2012ry}
  Y.~Grossman and D.~J.~Robinson,
  %``SU(3) Sum Rules for Charm Decay,''
  JHEP {\bf 1304} (2013) 067
  [arXiv:1211.3361 [hep-ph]].
  %%CITATION = ARXIV:1211.3361;%%
  %8 citations counted in INSPIRE as of 03 Jun 2014
  
  %\cite{Brod:2012ud}
\bibitem{Brod:2012ud}
  J.~Brod, Y.~Grossman, A.~L.~Kagan and J.~Zupan,
  %``A Consistent Picture for Large Penguins in D -> pi+ pi-, K+ K-,''
  JHEP {\bf 1210} (2012) 161
  [arXiv:1203.6659 [hep-ph]];
  see also J.~Brod, talk at this conference.
  %%CITATION = ARXIV:1203.6659;%%
  %63 citations counted in INSPIRE as of 03 Jun 2014
  
  %\cite{Bhattacharya:2014eca}
\bibitem{Bhattacharya:2014eca}
  B.~Bhattacharya, M.~Gronau, M.~Imbeault, D.~London and J.~L.~Rosner,
  %``Charmless B -> PPP Decays: the Fully-Symmetric Final State,''
  arXiv:1402.2909 [hep-ph].
  %%CITATION = ARXIV:1402.2909;%%
  %1 citations counted in INSPIRE as of 03 Jun 2014
  
  
%\cite{Chua:2013zga}
\bibitem{Chua:2013zga}
  C.-K.~Chua,
  %``Charmless Two-body Baryonic $B_{u,d,s}$ Decays Revisited,''
  Phys.\ Rev.\ D {\bf 89} (2014) 056003
  [arXiv:1312.2335 [hep-ph]].
  %%CITATION = ARXIV:1312.2335;%%

  
  %\cite{Jung:2012mp}
\bibitem{Jung:2012mp}
  M.~Jung,
  %``Determining weak phases from $B\to J/\psi P$ decays,''
  Phys.\ Rev.\ D {\bf 86} (2012) 053008
  [arXiv:1206.2050 [hep-ph]].
  %%CITATION = ARXIV:1206.2050;%%
  %12 citations counted in INSPIRE as of 03 Jun 2014


 %\cite{Ball:2004ye}
\bibitem{Ball:2004ye}
  P.~Ball and R.~Zwicky,
  %``New results on B ---> pi, K, eta decay formfactors from light-cone sum rules,''
  Phys.\ Rev.\ D {\bf 71} (2005) 014015
  [hep-ph/0406232];
  %%CITATION = HEP-PH/0406232;%%
  %482 citations counted in INSPIRE as of 03 Jun 2014
  %\cite{Ball:2004rg}
%\bibitem{Ball:2004rg}
%  P.~Ball and R.~Zwicky,
  %``B(D,S) ---> rho, omega, K*, phi decay form-factors from light-cone sum rules revisited,''
  Phys.\ Rev.\ D {\bf 71} (2005) 014029
  [hep-ph/0412079].
  %%CITATION = HEP-PH/0412079;%%
  %387 citations counted in INSPIRE as of 03 Jun 2014
  
  
  %\cite{Duplancic:2008ix}
\bibitem{Duplancic:2008ix}
  G.~Duplancic, A.~Khodjamirian, T.~Mannel, B.~Melic and N.~Offen,
  %``Light-cone sum rules for B ---> pi form factors revisited,''
  JHEP {\bf 0804} (2008) 014
  [arXiv:0801.1796 [hep-ph]].
  %%CITATION = ARXIV:0801.1796;%%
  %84 citations counted in INSPIRE as of 03 Jun 2014
  
  \bibitem{latticetalk}
  L.~Lellouch, talk at this conference;
  S.~Meinel, talk at this conference.
  
  
  
  %\cite{Braun:2003wx}
\bibitem{Braun:2003wx}
  V.~M.~Braun, D.~Y.~Ivanov and G.~P.~Korchemsky,
  %``The B meson distribution amplitude in QCD,''
  Phys.\ Rev.\ D {\bf 69} (2004) 034014
  [hep-ph/0309330].
  %%CITATION = HEP-PH/0309330;%%
  %121 citations counted in INSPIRE as of 03 Jun 2014

 %\cite{Bell:2013tfa}
\bibitem{Bell:2013tfa}
  G.~Bell, T.~Feldmann, Y.-M.~Wang and M.~W.~Y.~Yip,
  %``Light-Cone Distribution Amplitudes for Heavy-Quark Hadrons,''
  JHEP {\bf 1311} (2013) 191
  [arXiv:1308.6114 [hep-ph]].
  %%CITATION = ARXIV:1308.6114;%%
  %5 citations counted in INSPIRE as of 03 Jun 2014

  %\cite{Braun:2014vba}
\bibitem{Braun:2014vba}
  V.~M.~Braun and A.~N.~Manashov,
  %``Two-loop evolution equations for light-ray operators,''
  arXiv:1404.0863 [hep-ph].
  %%CITATION = ARXIV:1404.0863;%%
  
  
 %\cite{Feldmann:2014ika}
\bibitem{Feldmann:2014ika}
  T.~Feldmann, B.~O.~Lange and Y.~-M.~Wang,
  %``B-Meson Light-Cone Distribution Amplitude: Perturbative Constraints and Asymptotic Behaviour in Dual Space,''
  Phys.\ Rev.\ D {\bf 89} (2014) 114001
  [arXiv:1404.1343 [hep-ph]].
  %%CITATION = ARXIV:1404.1343;%%
 
  
  
  %\cite{Beneke:2011nf}
\bibitem{Beneke:2011nf}
  M.~Beneke and J.~Rohrwild,
  %``B meson distribution amplitude from B --> \gamma l \nu,''
  Eur.\ Phys.\ J.\ C {\bf 71} (2011) 1818
  [arXiv:1110.3228 [hep-ph]].
  %%CITATION = ARXIV:1110.3228;%%
  %6 citations counted in INSPIRE as of 03 Jun 2014
  
  %\cite{Braun:2012kp}
\bibitem{Braun:2012kp}
  V.~M.~Braun and A.~Khodjamirian,
  %``Soft contribution to $B\to \gamma \ell \nu_\ell$ and the $B$-meson distribution amplitude,''
  Phys.\ Lett.\ B {\bf 718} (2013) 1014
  [arXiv:1210.4453 [hep-ph]].
  %%CITATION = ARXIV:1210.4453;%%
  %4 citations counted in INSPIRE as of 03 Jun 2014
  
  
  %\cite{Wang:2013fya}
\bibitem{Wang:2013fya}
  K.~Wang and G.~Zhu,
  %``Flavor dependence of annihilation parameters in QCD factorization,''
  Phys.\ Rev.\ D {\bf 88} (2013) 014043
  [arXiv:1304.7438 [hep-ph]].
  %%CITATION = ARXIV:1304.7438;%%
  %1 citations counted in INSPIRE as of 03 Jun 2014

  
  %\cite{Chang:2012xv}
\bibitem{Chang:2012xv}
  Q.~Chang, X.-W.~Cui, L.~Han and Y.-D.~Yang,
  %``Revisiting the Annihilation Corrections in Non-leptonic $\bar{B}_s^0$ Decays within QCD Factorization,''
  Phys.\ Rev.\ D {\bf 86} (2012) 054016
  [arXiv:1205.4325 [hep-ph]].
  %%CITATION = ARXIV:1205.4325;%%
  %7 citations counted in INSPIRE as of 03 Jun 2014
  
  %\cite{Gronau:2012gs}
\bibitem{Gronau:2012gs}
  M.~Gronau, D.~London and J.~L.~Rosner,
  %``Rescattering Contributions to rare B-Meson Decays,''
  Phys.\ Rev.\ D {\bf 87} (2013) 3,  036008
  [arXiv:1211.5785 [hep-ph]].
  %%CITATION = ARXIV:1211.5785;%%
  %6 citations counted in INSPIRE as of 03 Jun 2014


  
  
\end{thebibliography}
\end{document}